# In-Memory Computing Architecture for Efficient Hardware Security

Hala Ajmi
*Electronics and Microelectronics Laboratory*
*Faculty of Sciences Of Monastir*
*University of Monastir*
Monastir, Tunisia
ajmihala@gmail.com

Fakhreddine Zayer
*Khalifa University of Science and Technology*
*Khalifa Univesity*
Abu Dhabi, UAE
Fakhreddine.Zayer@ku.ac.ae

Hamdi Belgacem
*Electronics and Microelectronics Laboratory*
*Faculty of Sciences Of Monastir*
*University of Monastir*
Monastir, Tunisia
Belgacemhamdi@gmail.com

*Abstract*—This paper presents an innovative approach utilizing in-memory computing (IMC) for the development and integration of AES (Advanced Encryption Standard) cipher technique. Our research aims to enhance cybersecurity measures for a wide range of applications for IoT, such as robotic self-driving and several uses contexts. Memristor (MR) design optimized for in-memory processing is introduced. Our work highlights the development of a 4-bit state memristor device tailored for various range of arithmetic functions in a hardware prototype of AES system. Additionally, we propose a pipeline AES design aimed at harnessing extensive parallelism and ensuring compatibility with MR devices. This approach enhances hardware performance by by managing larger data amounts, accelerating computational, and achieving greater precision demands. Compared to traditional AES hardware, AES-IMC demonstrates an approximate 30 % improvement in power with a comparable throughput rate. Compared with the latest AES-based NVM engines, AES-IMC achieves an impressive 62 % improvement in throughput at similar power dissipation levels. The IMC-developed design will protect against unintentional incidents involving unmanned devices, reducing the risks associated with hostile assaults such as hijacking and illegal control of robots. This helps to reduce the possible economic and financial losses caused by incidents.

*Index Terms*—AES cipher, memristive architecture, Hardware security, in-memory computing, FPGA implementation

## I. INTRODUCTION

The rapid evolution of computing, communication, and storage demands, fueled by the expansive growth of IoTs and the accumulation of vast datasets, poses significant challenges for conventional DRAM-based CMOS technology. In reaction, industries and researchers are rapidly looking into novel avenues, with nonvolatile memory (NVM) emerging as a promising frontier [1]. This concerted effort spans various domains, encompassing computing and storage [2], logic designs [3], as well as cutting-edge applications in neuromorphic computing, computer vision, [4] and oscillator circuits [5]. Illustrated in Figure 1, the capabilities of IMC (In-Memory Computing) provide effective approaches with tolerable switching latency, minimal power, and costs. [6].

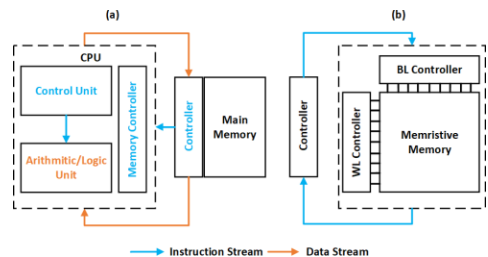

Fig. 1. Comparison between a) Von Neumann architecture and. b) In-Memory, computing paradigms

Despite their advantageous capabilities, NVMs are susceptible to integrity and security vulnerabilities [7]. Unlike traditional volatile memory, NVM retains data even after power-off, making it vulnerable to attacks by individuals with physical access to the system, who can readily scan memory contents and extract sensitive information [8] [9]. However, the protection of DRAM depends on its minimal retention period. [10]. Therefore, to safeguard data stored in NVMs, it is imperative to implement robust security mechanisms such as the Advanced Encryption Standard (AES), which offers a level of security comparable to DRAM's [11].

Implementing Real-time Memory Encryption (RTME) with a data stream cipher provides an effective solution to this vulnerability. In this approach, each cache memory line undergoes encryption or decryption before being read or written [12]. However, this enhanced security comes at a cost: there's a noticeable processing speed decrease occurs due to decryption latency, which acts as an additional cost for memory access. Furthermore, the processes of memory encryption (ME) and decryption (MD) entail significant energy overhead. Despite the advantages of the bulk Memory Encryption (ME) approach, such as eliminating runtime performance loss and reducing energy and task requirements for encryption, two significant objectives to overcome. Firstly, The process needs to be sufficiently rapid to decrease the vulnerability period when locked and respond instantly when unlocked. This becomes even more critical with the proliferation of multicore processors and the escalating demand for much larger main memory. Secondly, it necessitates energy-efficient encryption methods to accommodate the constraints of limited battery life.

In response to these challenges, we introduce AES-IMC [13], an innovative encryption architecture tailored for swift and energy-efficient NVM encryption. By capitalizing on the capabilities of in-memory computing, A substantial amount of multi-bit-line-level parallelism, low processing latency, and abundant internal memory bandwidth are all utilized by AES-IMC. With this arrangement, data transfer between the host and the memory is not necessary, thereby enhancing both speed and energy efficiency. This paper presents a series of significant contributions to the field. Firstly, We suggest using two 64-bit AES units of processing in our pipeline AES architecture, tailored to integrate with the AES-IMC design framework seamlessly. This design not only satisfies strict performance and energy efficiency requirements but also guarantees compatibility with MR crossbar topology while maintaining high parallel processing. Secondly, We put forward an AES pipeline with enhanced IMC features, which makes it easier to execute different AES encryption stages' arithmetic functions using IMC paradigms. Finally, we conduct a comprehensive comparative analysis, contrasting our proposed designs with conventional AES architectures and current AES-based NVM techniques. Through these contributions, we aim to advance the understanding and development of efficient and secure AES encryption methodologies based on in-memory computing (IMC).

## II. PROPOSED AES-IMC ARCHITECTURE

In this section, We describe in detail the suggested AES encryption technique, which entails executing a predetermined number of repetitions and applying a sequence of changes to the contents of the bits stream. The input is partitioned into two sub-units, each processing 64 bits. The encryption process leverages MR crossbar units for efficient computation of all transformations. Designed to optimize both throughput and power consumption, this architecture is tailored to meet the demands of various IoT applications.

### A. Memristor-based Moore Finite State Machine

Our work integrates memristive technology that is simulated and supports several levels of qualities, minimal delay, and low power use. [14]. This emulator is essential to our development process since it allows us to perform tests and debug the software of the AES encryption system even before hardware becomes available. The development schedule is accelerated by this initial verification, which also makes it easier to quickly identify and address possible problems. The application of pipeline strategies and IMC principles in similar or advanced cryptographic algorithms offers significant potential for enhancing security and achieving efficient, cost-effective processing.

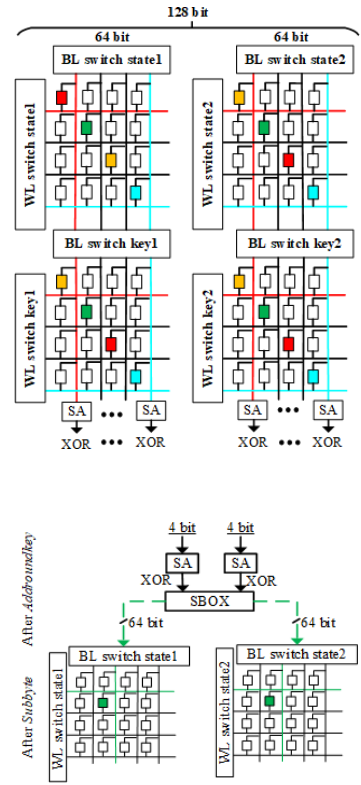

Fig. 3. *Subbyte* design of the AES-IMC.

### B. Addroundkey

In the Addroundkey step, the primary arithmetic operation involves designing an XOR function within the crossbar utilizing multi-state memristors. Accumulation is executed at the summing amplifier (SA) node. Figure 2 illustrates the addroundkey transformation process. Initially, the initial row of the data matrix is read into each SA's capacitor by activating the first-word line (red line) and selecting a column. The initial row of the key matrix is then read into the latch in each SA by activating the first word (red line) and selecting a column. The XOR outcome resulting from the two rows is then stored in each SA.

### C. Subbyte

Both of the input matrices of 4-bit portions are each decoded and fed into the S-box during the Subbyte transformation, as depicted in Figure 3. The results of the Addroundkey operation for the second row of the data matrix are latched in the Summing Amplifiers (SAs), and Subbyte processing is performed on each cell. The S-box functions as a lookup table, with the input byte directing to a specific output byte.

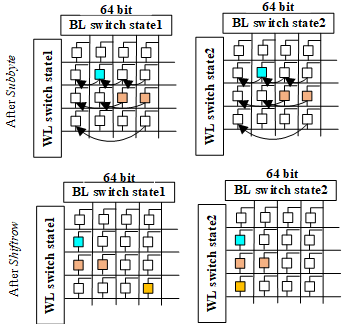

Fig. 4. *Shiftrow* design of the AES-IMC.

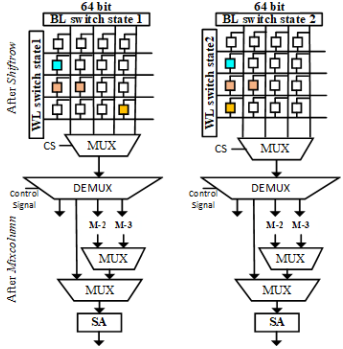

Fig. 5. *Mixcolumn* design of the AES-IMC.

### D. Shiftrow

The 4-bit output from each matrix's S-box needs to be written back to its beginning location after the Subbyte operation. To accomplish this, the column address and offset are combined, which moves the S-box output to a new location under the Shiftrow topology shown in Figure 4. This shifting procedure is aided by picking the first column with the control signal. Subsequently, after the Shiftrow phase, the entire bit is then stored in a single-bit latch until both Subbyte and Shiftrow operations are performed in the Summing Amplifiers (SAs). After being sent to the write driver, the values kept in the row buffer are then written back to the memory array. This row buffer serves to store intermediate results, thereby preventing unnecessary writes to the non-volatile memory rows.

### E. Mixcolumn

In the Mixcolumns stage, the data matrix undergoes multiplication by a predetermined matrix. A reversible linear adjustment is used to merge each column with its corresponding column found in the key, promoting diffusion in the encrypted output. Multiplication and XOR techniques are used to disassemble this transformation. Therefore, the Mixcolumns stage is decomposed into M-2 LUT and XOR operations. To hasten this transformation while keeping hardware overhead minimal, unoccupied memory rows are used as buffer rows for intermediate results.

The first phase in the Mixcolumns process is the M-2 conversion, which uses the same address decoding logic as the S-box and incorporates a MUX, as illustrated in Figure 5. Given that only one byte can be inputted to the Look-Up Table (LUT) at a time, this transformation occurs sequentially. To speed up this procedure, numerous M-2 LUTs are used to enable parallel processing. Various designs of multiplication-by-2 LUTs are evaluated, each offering different encryption speeds and overheads. Subsequently, the outputs are latched in a row buffer, and the data from this buffer is written to an available memory row. In the next phase, two rows are triggered to produce the XOR output of two cells in memory, which is written to an empty buffer row. Finally, the last step of the Mixcolumns entails determining the result of the Mixcolumns transformation.

## III. EXPERIMENTAL RESULTS

In this section, we present and analyze the results of the proposed design in comparison to existing works. Using Xilinx Vivado Integrated Design tools, the AES-IMC design has been assessed and synthesized. It has been implemented on the NEXYS 4 DDR FPGA board, which has an Artix-7 FPGA. A range of metrics are employed to evaluate the hardware design of AES-IMC on FPGA. Latency, throughput, and energy efficiency metrics are exploited. The maximum throughput (*Thr*) of the design implementation is influenced by a variety of parameters, including the maximum operating frequency ($F_{max}$), the latency cycles required to process a block ($L$), and the block size ($B_{size}$), energy efficiency is assessed through calculations of the total power dissipated by the architectures under consideration. This has the following mathematical expression:

$$Thr = \frac{F_{max} \times B_{size}}{L} \quad (1)$$

Low-resource implementations commonly do not use the maximum frequency, but limit it to a lower frequency, as is the case in RFID applications, i.e. FRF=13.56MHz. By having a common frequency, a reasonable comparison in regards to throughput per slot could be performed to evaluate all the architectures implemented in this paper. The throughput calculated at FRF is denoted (*Thr∗*) and calculated as follow:

$$Thr* = \frac{F_{RF} \times B_{size}}{L} \quad (2)$$

The calculation below shows the energy, *E*, that the implementation uses to execute only one block, where *P* is the total power dissipated.

$$E = \frac{P \times L}{F_{RF}} \quad (3)$$

The input size is 128 bits divided in two, each one of 64 bits, verifying that the complete encryption of a data packet is done in 26 cycles. The RTL design of the first round is visually depicted in Figure 6. The subsequent eight rounds follow a similar structure to the first round, except for the last round, which omits the MixColumns component. The floor plan of AES-IMC physical implementation into FPGA and the path of the different components of the first round are presented in Figure 7. Our algorithm requires 128 bits of input, which is split into two separate parts, each carrying 64 bits.

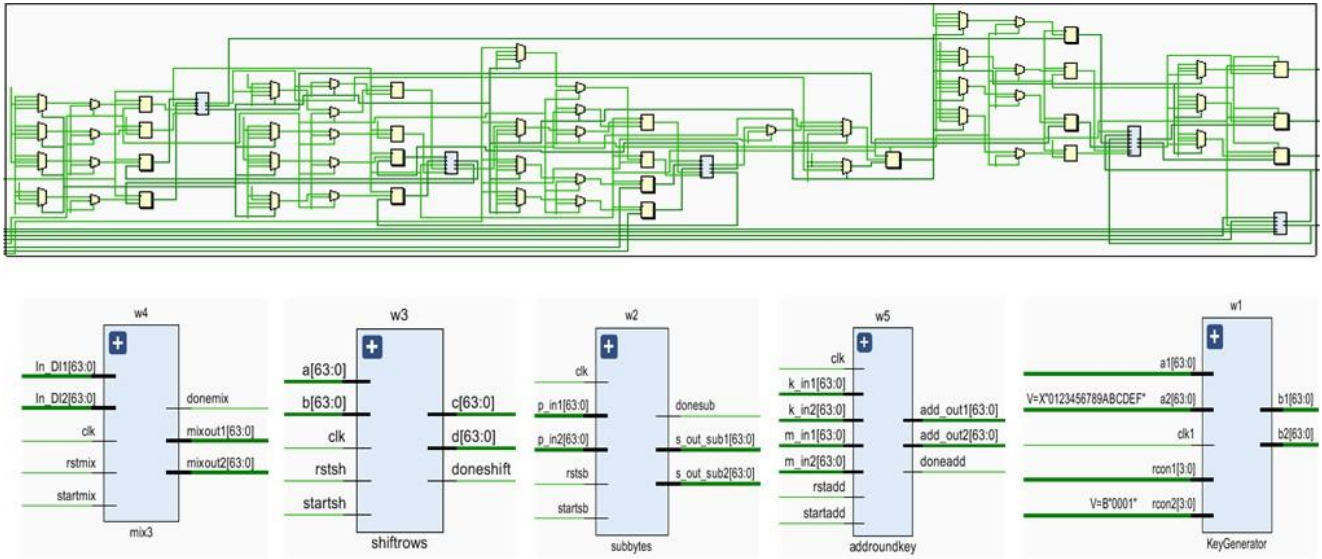

Fig. 6. RTL design of the first round of AES-IMC.

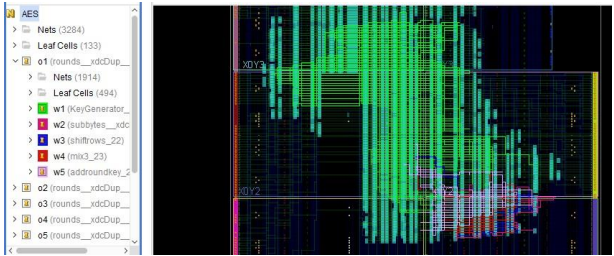

Fig. 7. Hardware implementation: Floor plan of AES-IMC, path of the different components of the first round.

Table I provides an extensive analysis of the resource consumption and performance of various topologies implemented on a range of FPGA devices. Each architecture's resource usage is explained in depth for the selected FPGA platforms that act as the implementation environment.

The findings suggest that implementations on the Spartan-6 FPGA exhibit lower slice counts and require fewer LUT elements compared to those on the Virtex FPGAs and our implementation. However, our proposed architecture, when deployed on LUT-4 FPGAs, demonstrates resource usage comparable to the iterative designs examined in the comparison.

In addition, Table 2 shows that the implemented AES-IMC has an outstanding minimal critical path of 0.806 ns, demonstrating higher efficiency when compared to other collected implementations.

Table 3 shows the findings of energy and power usage measurements among several FPGA devices. The report gives power estimations for each design based on a user-defined operational frequency of 13.56 MHz, which is consistent throughout each evaluation and at the standard working temperature. The current study was designed to evaluate energy usage, hence only power outputs were taken into account. Static power is consistent over various configurations of the same FPGA board, whereas dynamic power fluctuates according to the circuit's changing activity.

The present study compares AES-IMC implementation with various solutions that depend on technology at the system degree. Every computational structure is designed to encrypt incoming streams of information concurrently while remaining within a limited area limitation, increasing the number of AES units for each platform. Table 4 shows system setups for different hardware platforms. Among all hardware implementations, AES-IMC boasts the maximum throughput with a Data Processing Rate (DPR) of 445 GB/s. The suggested design requires less space per cipher, allowing more ciphers to run concurrently within an identical space limit, resulting in a larger DPR. This improvement in DPR is also obtained by lowering the critical phase's latency. Furthermore, the AES-IMC computing platforms' multi-bit in-memory encryption and non-volatility lead to improved energy efficiency, evaluated at 7.03 pJ/bit, which outperforms that of leading platforms. The suggested architecture has a benefit in energy efficiency over ASIC implementations due to its efficient in-memory computing and lower energy usage per encryption operation.

## IV. CONCLUSION

The present research proposes a novel hardware tool for AES encryption that focuses on increasing processing efficiency while exploring MR architecture. The proposed AES-IMC method encrypts several blocks of memory within the main memory at the same time but does not expose outcomes to the memory bus. Experimental results demonstrate AES-IMC's advantages over conventional AES engines, including higher throughput, faster data processing speeds, and lower energy usage.

TABLE I
UTILIZATION OF RESOURCES AND PERFORMANCE RESULTS FOR VARIOUS AES DESIGNS

| implementation | FPGA Platform | input (bit) | Key (bit) | FF | LUT | SLC | $F_{max}$ (MHz) | L (Cycles) | Thr (Mbps) | Thr*(Mbps) |
|---|---|---|---|---|---|---|---|---|---|---|
| [15] | Artix7 XC7A200T | 128 | 128 | 2911 | 1512 | 359 | 311.72 | 59 | 676.276 | 29.41 |
| [16] | Virtex II | 128 | 128 | 271 | 1862 | 976 | 60.94 | **26** | 300.01 | **66.76** |
| [17] | Virtex IV XC4VLX200 | 128 | 128 | 684 | 2127 | 1120 | 112.37 | 1000 | 14.383 | 1.73 |
| [16] | Virtex V | 128 | 128 | 271 | 1391 | 456 | 96.04 | **26** | 472.81 | **66.76** |
| [15] | Virtex7 XC7VX90T | 128 | 128 | 1817 | 1271 | 551 | 308.64 | 59 | 669.59 | 29.41 |
| [18] | XCZU9EG | 128 | 128 | 4296 | 15029 | 3262 | 220 | **10** | 2816 | **173.56** |
| [19] | Spartan 7 XC7S75FGGA484-1 | 128 | 128 | 1900 | 1425 | 430 | 213.7 | 57 | 480 | 30.45 |
| **AES-IMC** | Xc7A100T-CSG324 | 128 | 128 | 1245 | 2836 | 468 | 108.9 | **26** | 536.12 | **66.76** |

TABLE II
THE COMPARISON OF THE CRITICAL PATH OF OTHER RELATED WORK TO THE PRESENTED DESIGN.

| work | FPGA Platform | Path delay(ns) |
|---|---|---|
| [20] | XC7VX690T | 1.6 |
| [21] | XC5VLX30 | 3.6 |
| [22] | XC5VLX330 | 3.13 |
| [23] | XC5VLX85 | 1.56 |
| [24] | XC7VX690T | 1.23 |
| **AES-IMC** | Xc7A100T-CSG324 | 0.806 |

TABLE III
ENERGY AND POWER CONSUMPTION MEASUREMENTS FOR VARIOUS FPGA IMPLEMENTATIONS.

| Work | FPGA platform | input (bit) | Key (bit) | P (mW) | $E^*(\mu J)$ | $E^*$/ (bit) |
|---|---|---|---|---|---|---|
| [15] | Artix7(XC7A200T) | 128 | 128 | 0.184 | 0.80 | 6.25 |
| [17] | Virtex IV XC4VLX200 | 128 | 128 | 0.261 | 19.247 | 150.37 |
| [15] | Virtex7(XC7VX90T) | 128 | 128 | 0.463 | 2.01 | 15.7 |
| [18] | XCZU9EG | 128 | 128 | 1.17 | 0.86 | 6.74 |
| [19] | Spartan 7 xc7s75fgga484-1 | 128 | 128 | 563.5 | 2.368 | 18.5 |
| **AES-IMC** | Xc7A100T-CSG324 | 128 | 128 | **0.098** | **0.18** | **1.406** |

TABLE IV
COMPARISON OF PERFORMANCE OF VARIOUS AES-128BIT TECHNOLOGICAL ARCHITECTURES AT FMAX = 30MHZ.

| implementation | Area ($\mu m^2$) | L (Cycles) | P (W) | E (nJ) | Thr(Mbps) |
|---|---|---|---|---|---|
| CMOS ASIC [25] | 4400 | 336 | 0.013 | 6.6 | 5.16 |
| Memristive CMOL [26] | 320 | 470 | 0.309 | 10.3 | 3.69 |
| DW-AES Baseline [27] | 78 | 1022 | 0.072 | 2.4 | 1.71 |
| DW-AES Pipeline [27] | 83 | 2652 | 0.069 | 2.3 | 0.65 |
| DW-AES Multi-issue [27] | 155 | 1320 | 0.081 | 2.7 | 1.31 |
| **AES-IMC** | ~ 83 | **26** | 0.098 | **0.9** | 147.6 |